\documentclass[prl,twocolumn,preprintnumbers,amsmath,amssymb,superscriptaddress]{revtex4}
\usepackage{graphicx}  
\usepackage{dcolumn}   
\usepackage{bm}        
\usepackage{bbm}       
\usepackage{amsmath, amssymb, graphics}

\newcommand{\bra}[1]{\langle#1\vert} 
\newcommand{\ket}[1]{\vert#1\rangle} 
\newcommand{\beq}{\begin{equation}}
\newcommand{\eeq}{\end{equation}}
\newcommand{\bea}[1]{\begin{equation}\begin{array}{#1}}
\newcommand{\eea}{\end{array}\end{equation}}
\newcommand{\beqn}{\begin{eqnarray}}
\newcommand{\eeqn}{\end{eqnarray}}
\DeclareMathOperator{\tr}{Tr}  
\providecommand{\openone}{\mathbbm{1}}

\usepackage{color}
\definecolor{red}{rgb}{0.9,0.0,0.1}

\begin{document}

\title{A Factorization Law for Entanglement Decay}

\author{Thomas Konrad}
\email{konradt@ukzn.ac.za}
\affiliation{Quantum Research Group, School of Physics, University of KwaZulu-Natal, Private Bag X54001, Durban 4000, South Africa}
\author{Fernando de Melo}
\email{fmelo@pks.mpg.de}
\affiliation{Max-Planck-Institut f\"ur Physik komplexer Systeme, N\"othnitzer Str.38, D-01187 Dresden, Germany}
\author{Markus Tiersch}
\email{mtiersch@pks.mpg.de}
 \affiliation{Max-Planck-Institut f\"ur Physik komplexer Systeme, N\"othnitzer Str.38, D-01187 Dresden, Germany}
 \author{Christian Kasztelan}
 \email{kasztelan@physik.rwth-aachen.de}
 \affiliation{Institut f\"ur Theoretische Physik C, RWTH Aachen, D-52056 Aachen, Germany}
 \author{Adriano Arag\~ao}
 \email{aharagao@if.ufrj.br}
 \affiliation{Instituto de F\'{\i}sica, Universidade Federal do Rio de Janeiro, Caixa Postal 68.528, CEP 21945-970, Rio de Janeiro, RJ, Brazil}
 \affiliation{Max-Planck-Institut f\"ur Physik komplexer Systeme, N\"othnitzer Str.38, D-01187 Dresden, Germany}
\author{Andreas Buchleitner}
\email{abu@pks.mpg.de} 
\affiliation{Max-Planck-Institut f\"ur Physik komplexer Systeme, N\"othnitzer Str.38, D-01187 Dresden, Germany}



\begin{abstract}

 We present a simple and general factorization law for 
quantum systems shared by two parties, which describes the time evolution of
entanglement upon passage of either component through an arbitrary
noisy channel.
The robustness of entanglement-based quantum
information processing protocols is thus easily and fully
characterized by a single quantity.
\end{abstract}

 \maketitle


Whenever we contemplate the potential technological applications of quantum
information theory~\cite{bennett:247}, from secure quantum communication over
quantum 
teleportation~\cite{bennett:1895} to quantum com\-pu\-ta\-tion~\cite{nielsen},
we need to worry about the unavoidable and detrimental coupling of
any such quantum device to uncontrolled degrees of
freedom -- typically lumped together under the label
``environment''. Environment coupling induces
decoherence~\cite{schlosshauer:1267,zurek:715,joos}, i.e., it gradually
destroys the phase relationship between quantum states, and thus their ability
to interfere. In composite quantum systems, these phase relationships (or
``coherences'') are at
the origin of strong quantum correlations between measurements on distinct
system constituents -- which then are \emph{entangled}. The promises of quantum
information technology rely on exploring precisely these non-classical 
correlations.

\begin{figure}
\begin{center}
\includegraphics{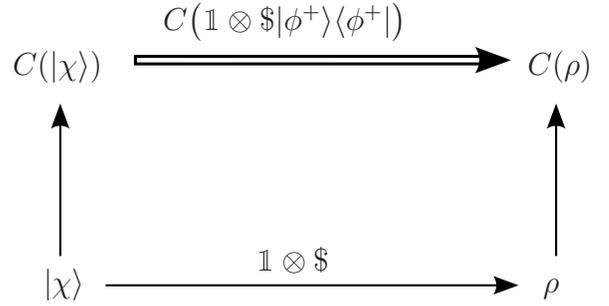}
\end{center}
\caption{
  From state evolution to entanglement dynamics.
  Hitherto, the time evolution of entanglement $C$ under open system dynamics
  had to be deduced from the time evolution of the state $\ket{\chi}$.
  However, it turns out that knowlegde of the entanglement evolution of
  the maximally entangled state $\ket{\phi^+}$ under
  the channel $\openone\otimes\$$ suffices
  to establish a direct mapping of $C(\ket{\chi})$ onto $C(\rho)$,
  without detour over the state.
}
\label{fig1}
\end{figure}

Yet, 
entanglement is \emph{not} equivalent to many-particle coherences: it is an
even stronger property, and hard to quantify -- all commonly
accepted entanglement measures~\cite{mintert:207, horodecki-2007} are
\emph{nonlinear} functions of the density 
matrix which describes the state of the composite quantum system, and in
particular the coherences.
While an elaborate theory on the time evolution of quantum states under
environment coupling is at our hands, virtually no general results on
entanglement dynamics have been stated.
%
Hitherto, the time evolution of
entanglement always needed to be \emph{deduced} from the time evolution of the
state~\cite{karol,dodd:052105,roos:220402,carvalho:230501,yu:140403,santos:040305,dur:180403,marcelo,carvalho:190501,ebc}.
In the present Letter, a {\em direct} relationship between
the initial and final entanglement of an arbitrary bipartite
state of two qubits (basic units of quantum information) subject to incoherent
dynamics in one system component is derived,
which, as illustrated in Fig.~\ref{fig1},
renders the solution of the corresponding state evolution equation
obsolete.
Our result can be directly applied to input/output processes, such
as gates used in sequential quantum computing. Moreover, it allows to infer the
evolution of entanglement under certain time-continuous influences
of the environment, e.g. phase- and amplitude damping.

Let us consider entangled states of qubit pairs, with one qubit being
subject to an arbitrary channel $\$$ -- which may represent the influence
of an environment, of a measurement, or of both. In order to illustrate the
situation, we consider  a source which emits a particle to the left and
another one to the right. Each particle on its own carries one qubit 
of quantum information (in general a superposition or mixture of two
basis states $\ket{0}$ and $\ket{1}$).  We therefore also refer to the
particles as ``left'' and ``right''
qubit. Let the particles leaving the source be
in a pure state $\ket{\chi}$:
\beq
\ket{\chi}=\sqrt{\omega}\ket{00}+\sqrt{1-\omega}\ket{11} \; , 
\label{init}
\eeq
with $0 \le \omega \le 1$, i.e.,
for values of $\omega$ between zero and one
the particles are in a coherent superposition
of both being in state $\ket{0}$ and both being in state $\ket{1}$.
Any pure state can be written in this
form, modulo local unitary operations. Since our
results are not affected by these local unitaries, as we will see
below, this choice of the pure initial state does not impose a restriction
of generality.

We quantify entanglement 
by concurrence $C$~\cite{wootters:2245}, which implies for the state
$\ket{\chi}$: 
$C(\ket{\chi})=2\sqrt{\omega (1-\omega)}$.
For $\omega$ in \eqref{init} equal to zero or one,  the state's
entanglement and hence its concurrence vanishes; 
$\omega=1/2$ implies 
$\ket{\chi}=\ket{\phi^+}$, one of the maximally entangled Bell states,
with maximal concurrence one.

\begin{figure}
\begin{center}
\begin{tabular}{c}
\includegraphics[scale=0.4]{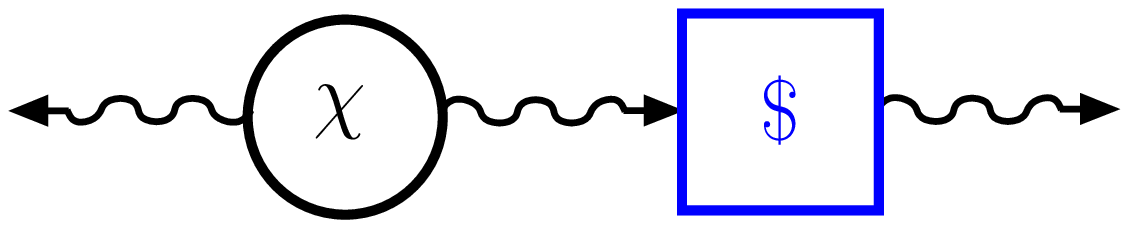}\\
a)\\
\\ \\
\includegraphics[scale=0.4]{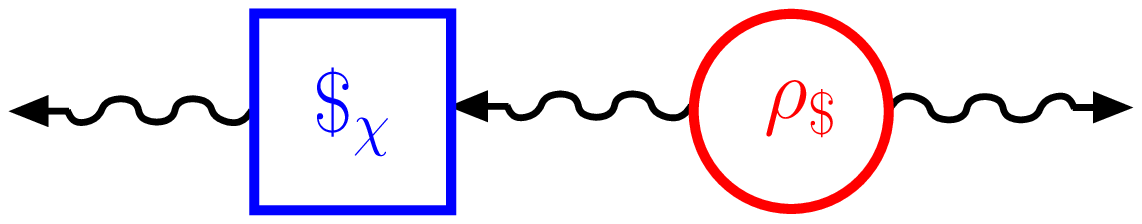}\\
b)
\end{tabular}
\end{center}
\caption{Entanglement decay scenarios. a) Laboratory scenario:  one of the
  qubits of the initial state $\ket{\chi}$, the ``right'' one, undergoes the
  action of a general quantum channel $\$$.  b) Dual scenario:  the same end
  result is obtained by interchanging the role
  of states and channels: the ``left'' qubit of a mixed state  $\rho_\$$
  undergoes the action of the quantum channel $\$_\chi$. 
  Circles represent sources of entangled states, squares symbolize channels.}
\label{fig2}
\end{figure}

Now the right qubit traverses an arbitrary quantum channel $\$$, as
illustrated in Fig.~\ref{fig2}a), and we want to derive the qubits'
entanglement 
hereafter.
To do so, note that the qubits' final
state must be the same as in a dual picture~\cite{jamiolkowski},  where the
roles of 
initial state and channel are interchanged, 
as depicted in Fig.~\ref{fig2}b).
Thus, the two-qubit state $\ket{\chi}$ is identified with 
a qubit channel $\$_\chi$, and
the qubit channel $\$$ with a two-qubit state $\rho_\$$;
symbolically:  
\beq
\frac{\left (\openone\otimes\$\right )\ket{\chi}\bra{\chi}}{p^\prime}=
\frac{\left ( 
  \$_\chi\otimes\openone\right ) \rho_\$}{p}\; . 
\label{dual}
\eeq 
Here, $p^\prime = \tr [ \left (\openone\otimes\$\right )\ket{\chi}\bra{\chi}]$
and $p = \tr [\left ( 
\$_\chi\otimes\openone\right ) \rho_\$]$ are the probabilities for channels
$\$$ and $\$_\chi$ to act 
on the states $\ket{\chi}\bra{\chi}$
and 
$\rho_\$$, respectively. Thus, we also account for non-trace-preserving
channels, where the particle number is not conserved. 

\begin{figure}
\begin{center}
\includegraphics[scale=0.4]{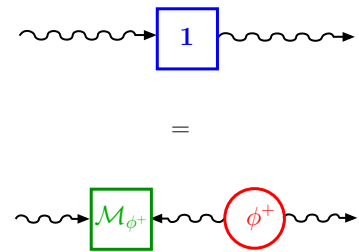}
\caption{Quantum Teleportation identity. The teleportation protocol transfers
  the incoming state from the left to the outgoing state on the
  right~\cite{note0}. The procedure consists of a composite Bell measurement
  on the incoming qubit and on one qubit of an auxiliary,
  maximally entangled state $\ket{\phi^+}$. We restrict to the case where
  the measurement results in a projection ${\cal M}_{\phi^+}$ on
 the state $\ket{\phi^+}$.
}
\label{fig3}
\end{center}
\end{figure}

\begin{figure}
\begin{center}
\includegraphics[scale=0.35]{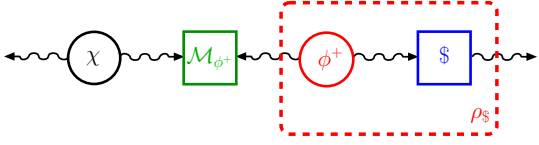}
\caption{
  Building in the teleportation identity.
  The right qubit of $\ket{\chi}$ undergoes the action of
  a quantum channel $\$$, after intermediate teleportation.
  The maximally entangled state $\ket{\phi^+}$ together with
  the action of the channel $\openone\otimes\$$
  yield the source of the mixed state $\rho_\$$.
}
\label{fig4}
\end{center}
\end{figure}

\begin{figure} 
\begin{center} 
\includegraphics[scale=0.4]{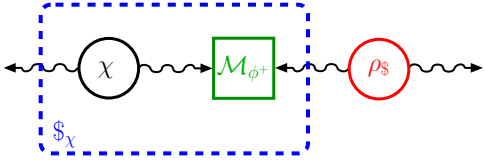} 
\caption{
  Entanglement swapping.
  Transfer of entanglement between the qubit pairs of $\ket{\chi}$ and
  $\rho_\$$, respectively, to entanglement between the outgoing pair of qubits.
  The Bell measurement ${\cal M}_{\phi^+}$ together with the source of
  the entangled state $\ket{\chi}$ constitute the quantum channel $\$_\chi$
  for the left qubit of state $\rho_\$$.
}
\label{fig5}
\end{center}
\end{figure}

We now need to determine $\$_\chi$ and $\rho_\$$ explicitely.
For this purpose, we first remember
that Quantum Teleportation~\cite{bennett:1895} is a means to transfer the
state of one system to another one, in principle with perfect
fidelity. Consequently, teleporting the right qubit of the state $\ket{\chi}$
assisted by the maximally entangled state $\ket{\phi^+}$ leaves the state
$\ket{\chi}$ invariant. This invariance is depicted in Fig.~\ref{fig3}.
We therefore obtain the same final state as in
the situation we considered so far (Fig.~\ref{fig2}a)) -- if we
replace the source preparing state $\ket{\chi}$ by a source which
prepares $\ket{\chi}$ followed by a teleportation of the right
qubit as shown in Fig.~\ref{fig4}. 
Let us now consider the source of the
qubit pair in state $\ket{\phi^+}$, which we inserted with the
teleportation. The succession of processes
  influencing the left qubit and those acting on 
the right qubit of the pair can be altered without
consequences for the final state. For this reason we can replace the
source producing the state $\ket{\phi^+}$
together with the channel $\$$ acting on the right qubit by
yet another source which immediately prepares the state 
\beq
\rho_\$:=(\openone\otimes\$)\ket{\phi^+}\bra{\phi^+}/p^{\prime\prime}\,,
\eeq
where $p^{\prime\prime} = \tr
[(\openone\otimes\$)\ket{\phi^+}\bra{\phi^+}]$, see
Fig.~\ref{fig4}.
The resulting scheme in Fig.~\ref{fig5} transfers entanglement
between the qubit pairs prepared in states $\ket{\chi}$ and
$\rho_\$$ to entanglement between the left qubit of
the first pair and the right qubit of the second pair. This scheme is called
entanglement swapping~\cite{bennett:1895,zukowski:4287,pan:3891}.

Finally, we define $\$_\chi$ to be the channel
corresponding to the change of the left qubit of
$\rho_\$$ in Fig.~\ref{fig5}, which includes a projection ${\cal
  M}_{\phi^+}$ of the left qubit of state $\rho_\$$ and the right qubit of
  $\ket{\chi}$ on $\ket{\phi^+}$. Channel  $\$_\chi$ can be
  interpreted as imperfect teleportation assisted by state $\ket{\chi}$, 
leaves
the resulting state in general non-normalized, and
can be expressed in the particularly simple form:
\beq
\left (\$_\chi\otimes\openone\right ) \rho_\$= \left
  (M\otimes\openone\right ) \rho_\$
\left ( M^\dag\otimes\openone\right ) \, ,
\label{kraus}
\eeq 
with $M = \big( \sqrt{\omega}\ket{0}\bra{0}+
\sqrt{1-\omega}\ket{1}\bra{1} \big) / \sqrt{2}$.
The normalized final state
$(\$_\chi\otimes\openone)\rho_\$/p$ is the same as
$(\openone\otimes\$)\ket{\chi}\bra{\chi}/p^\prime$, as spelled out by 
\eqref{dual} and in Fig.~\ref{fig2},
but the  entanglement evolution induced by 
the particular channel $\$_\chi$ can be 
deduced more easily, as we will now demonstrate.
The concurrence $C$ of the final state
$\rho^\prime=\left(\openone\otimes\$\right)\ket{\chi}\bra{\chi}/p^\prime$
is given by 
\beq
C(\rho^\prime)=\max\left\{0,\sqrt{\xi_1}-\sqrt{\xi_2}-\sqrt{\xi_3}-\sqrt{\xi_4}\right\}   
\; ,   
\label{conc}
\eeq
where the $\xi_i$ are the eigenvalues of the matrix
$\rho^\prime\cdot{\tilde\rho^\prime}$,
in decreasing order, with
${\tilde\rho^\prime}=(\sigma_y\otimes\sigma_y)\cdot\rho^{\prime *}\cdot(\sigma_y\otimes\sigma_y)$, and
$\rho^{\prime *}$
the complex conjugate of $\rho^\prime$, 
in the canonical basis. In order to evaluate this
expression, we use relation~\eqref{dual} together with \eqref{kraus}.
We write explicitly: 
\begin{equation}
\rho^\prime\cdot\tilde{\rho^\prime} =\frac{1}{p^2}\left(M\otimes\openone\right
  )\rho_\$\cdot\left 
  [M\sigma_yM\otimes\sigma_y\right ]\cdot\rho_\$^*\cdot\left
  [M\sigma_y\otimes\sigma_y\right ]\, ,
\end{equation}
where we employed that $M=M^\dagger=M^*$. For invertable $M$ ~\cite{note1},
it follows that the eigenvalues of
$\rho^\prime\cdot\tilde{\rho^\prime}$ and
$\rho_\$\cdot\tilde{\rho_\$}$  are proportional, since 
\bea{ccl}
& & \det \left[ \rho^\prime\cdot\tilde{\rho^\prime}-\xi\openone \right] \\
& = &
\det \left[ \left( M\otimes\openone \right)^{-1} \right]
\det \left[ M\otimes\openone \right]
\det \left[ \rho^\prime\cdot\tilde{\rho^\prime}-\xi\openone \right] \\
& = &
\det \left[
\left( M\otimes\openone \right)^{-1}
\rho^\prime\cdot\tilde{\rho^\prime}
\left( M\otimes\openone \right)-\xi\openone
\right] \\
& = &
\left [ \frac{1}{4 p^2}\omega(1-\omega)\right]^4 
\det \left[ \rho_\$ \cdot \tilde{\rho_\$}-\mu\openone \right]
\, ,
\eea
where $\mu = \xi \left(\omega(1-\omega)/4 p^2\right)^{-1}$, and we used $M\sigma_y M= \sqrt{\omega(1-\omega)}\sigma_y/2$ in
order to obtain the last equality.
Eq.~\eqref{conc}, together with
the
definitions of
$\rho_\$=(\openone\otimes\$)\ket{\phi^+}\bra{\phi^+}/p^{\prime\prime}$,  
$\rho^\prime=(\openone\otimes\$)\ket{\chi}\bra{\chi}/p^\prime$,  and
$C(\ket{\chi}) = 2 \sqrt{\omega(1-\omega)}$, thus lead~\cite{note2} to our
central result:  
\beq
C \left[ \left( \openone\otimes\$ \right) \ket{\chi}\bra{\chi} \right]
=
C \left[ \left( \openone\otimes\$ \right) \ket{\phi^+}\bra{\phi^+} \right]\,
C(\ket{\chi})
\label{ratio}
\eeq
-- the entanglement reduction under a one-sided noisy channel is
\emph{independent} of the initial state $\ket{\chi}$ and
\emph{completely determined} by the channel's action on the maximally entangled
state.
Thus, if we know the time evolution of the Bell state's
entanglement, we know it for \emph{any} pure initial state~\cite{note3}. 
This result can also be interpreted in terms of entanglement swapping
between a pure state $\ket{\chi}$ and a mixed state~$\rho_\$$, 
leading to the final state~$\rho^\prime$, due to the equivalence of the 
processes represented in Figs.~\ref{fig2} and~\ref{fig5}.


The factorization law (\ref{ratio}) can be generalized for mixed initial
states $\rho_0$, by virtue of the convexity of entanglement monotones such as
concurrence, and given an optimal pure state decomposition
$\rho_0=\sum_jp_j\ket{\psi_j}\bra{\psi_j}$, in the sense that the average
concurrence over this pure state decomposition is minimal~\cite{bennett:3824}.
It then immediately follows, by convexity, that
$
C \left[ \left( \openone\otimes\$\right) \rho_0 \right]
=
C \left[ \sum_j p_j \left(\openone\otimes\$\right)
  \ket{\psi_j}\bra{\psi_j} \right]
\leq
\sum_j p_j C \left[ \left(\openone\otimes\$\right)
  \ket{\psi_j}\bra{\psi_j}\right ]
$, and
application of (\ref{ratio}) leaves us with 
\begin{equation}
C \left[ \left( \openone\otimes\$ \right) \rho_0 \right]
\leq
C \left[ \left( \openone\otimes\$ \right)
  \ket{\phi^+}\bra{\phi^+} \right]
\, C(\rho_0)
\;.
\label{fact_mixed}
\end{equation}
This inequality holds for all one-sided channels $\$$, and has an immediate
generalization for local two-sided channels
$\$_1\otimes\$_2=(\$_1\otimes\openone)(\openone\otimes\$_2)$: 
%
\begin{align}
\label{two-side}
C \left[ \left( \$_1\otimes\$_2 \right) \rho_0 \right]
\leq&
\, C \left[ \left( \$_1\otimes\openone \right) \ket{\phi^+}\bra{\phi^+} \right]\\
 &\times \, C \left[ \left( \openone\otimes\$_2 \right) \ket{\phi^+}\bra{\phi^+} \right]
C(\rho_0)
\;.\nonumber
\end{align}
The concurrence after passage through a two-sided channel is thus bounded from
above, which immediately implies a sufficient criterion for finite-time
disentanglement~\cite{karol,dodd:052105,yu:140403} 
of {\em arbitrary initial states}, in terms of the evolution
of the concurrence of the maximally entangled state under either one of the
one-sided channels (e.g., choose $\$_1$ or $\$_2$ induced by infinite
temperature or depolarizing environments).   

Let us finally identify relevant cases when 
equality in \eqref{fact_mixed} holds. For that purpose,
we consider mixed states that are obtained after the application of
a 
one-sided channel to an arbitrary pure state,
$\rho_0 = (\openone\otimes\$) \ket{\psi_0}\bra{\psi_0}$.
This occurs, for instance, if the qubit originally 
prepared in a pure state
suffers amplitude decay, and the resulting mixed state again is
subject to decay dynamics. This is tantamount to  
the concatenation of channels on one side,
$\left ( \openone\otimes\$_2\right )\left (\openone\otimes\$_1\right )
\ket{\psi_0}\bra{\psi_0}$,
what   
can be lumped together 
as one channel
which combines both actions,
$\left ( \openone\otimes\$_2\right ) \left (\openone\otimes\$_1 \right
) = \openone\otimes\$_{2,1} $.
In a similar vein as for \eqref{two-side}, and
using the factorization relation~\eqref{ratio} for pure states,
we deduce 
%
\begin{align}
\label{concatenation}
C \left[ \left( \openone\otimes\$_{2,1} \right) 
         \ket{\phi^+}\bra{\phi^+} \right] 
\leq &\,C \left[ ( \openone\otimes\$_2 ) \ket{\phi^+}\bra{\phi^+} \right]\\ 
&\times \, C \left[ ( \openone\otimes\$_1 ) \ket{\phi^+}\bra{\phi^+} \right] \,.\nonumber
\end{align}
%
The initial state's concurrence rescales both sides of the equation
by the same amount, and therefore it is omitted in above equation. It is now sufficient to
investigate the time dependence of the maximally entangled state's concurrence
under the concatenated channels (much as for the evaluation of (\ref{ratio})):
if all of these are of the form 
$C(t)=\exp(-\Gamma t)$ (which is the case, e.g., for $\$_1$ an amplitude decay
and $\$_2$ a dephasing channel), 
then equality holds in (\ref{fact_mixed}), with
$\$=\$_1$, $\rho_0 = (\openone\otimes\$_2)
\ket{\psi_0}\bra{\psi_0}$, and $C(\rho_t)=\exp(-\Gamma_1t) C(\rho_0)$ (and,
equivalently, for the roles of channels 1 and 2 interchanged).   

In conclusion,
equations~(\ref{ratio},\ref{fact_mixed},\ref{two-side},\ref{concatenation}) 
provide us with the first closed expression for
the time evolution of a bipartite entangled state
under general local, single- and two-sided channels,
without recourse to the time evolution of the underlying quantum state itself.
This is a general result inherited from
the Jamio{\l}kowski isomorphism~\cite{jamiolkowski}, which 
is here ``lifted'' {\em from state to entanglement evolution}
(see Fig.~\ref{fig1}), and
eases the experimental characterization of entanglement  dynamics under
unknown channels dramatically: instead of exploring the time-dependent action
of the channel on {\em all} initial states, it suffices to probe the
entanglement evolution of the maximally entangled state alone.

\begin{acknowledgments}
 We thank  J.~Audretsch, H.-P.~Breuer, L.~Davidovich, L.~Diosi and C.~Viviescas for
 stimulating discussions. Partial support by DAAD/CAPES under a PROBRAL grant
 is gratefully acknowledged. 
 \end{acknowledgments}


\begin{thebibliography}{10}
\bibitem{note0}
Here only the measurement of $\ket{\phi^+}$ is used. The remaining three
  possible measurement outcomes would imply local operations on the outgoing
  qubit, which can be incorporated in the action of the channel $\$$.
\bibitem{note1}
This can be assumed without loss of generality, since otherwise, for $\omega
  =0$ or $\omega =1$, and hence $M\sigma_yM=0$, concurrence vanishes.
\bibitem{note2}
In this last equation we used that $C(\alpha \rho)=\alpha C(\rho)$ for positive
  real $\alpha$ and $ p^\prime = 4 p\, p^{\prime\prime}$. The latter can be
  derived straightforwardly, by careful comparison of the individual
  normalization factors. In order to compute the entanglement of the normalized
  final state $C \left[ \left( \openone\otimes\$ \right) \ket{\chi}\bra{\chi}
  \right]$ has to be devided by $p^\prime$.
\bibitem{note3}
Note that our choice (\ref{init}) of $\ket{\chi}$ does not restrict the
  generality of relation (\ref{ratio}). An initial state
  $\ket{\chi^\prime}\bra{\chi^\prime}= \left({\mathcal U_a}\otimes {\mathcal
  U_b}\right) \ket{\chi}\bra{\chi}$ leads to the same result (\ref{ratio})
  since unitary operation $\mathcal U_a$ does not affect the concurrence of the
  final state and unitary operation $\mathcal U_b$ can be incorporated into the
  channel: $C \left[ \left( \openone\otimes\$ \right)
  \ket{\chi^\prime}\bra{\chi^\prime} \right]= C \left[ \left(
  \openone\otimes\$\,{\mathcal U_b} \right) \ket{\chi}\bra{\chi} \right]=C
  \left[ \left( \openone\otimes\$ \,{\mathcal U_b}\right)
  \ket{\phi^+}\bra{\phi^+} \right]\, C(\ket{\chi})= C \left[ \left(
  \openone\otimes\$ \right) \ket{\phi^+}\bra{\phi^+} \right]\,
  C(\ket{\chi^\prime})$.
\end{thebibliography}

\end{document}